\documentclass[prl,preprint,groupedaddress,showpacs]{revtex4-1}

\usepackage{amsmath}
\usepackage{amssymb}
\usepackage{amsfonts}
\usepackage{mathrsfs}
\usepackage{epsfig}
\usepackage{bm}

\begin{document}

\title{Nonlinear laser driven donut wakefields for positron and electron acceleration}
\author{J. Vieira}
\email{jorge.vieira@ist.utl.pt}
\author{J.T. Mendon\c ca}
\email{titomend@ist.utl.pt}

\affiliation{GoLP/Instituto de Plasmas e Fus\~{a}o Nuclear-Laborat\'orio Associado,  Instituto Superior T\'{e}cnico, Universidade de Lisboa, Lisboa, Portugal}

\begin{abstract}
We show analytically and through three-dimensional particle-in-cell simulations that non-linear wakefields driven by Laguerre-Gaussian laser pulses can lead to hollow electron self-injection and positron acceleration. We find that higher order lasers can drive donut shaped blowout wakefields with strong positron accelerating gradients comparable to those of a spherical bubble. Corresponding positron focusing forces can be more than an order of magnitude stronger than electron focusing forces in a spherical bubble. Required laser intensities and energies to reach the non-linear donut shaped blowout are within state-of-the-art experimental conditions.
\end{abstract}

\maketitle

Laser plasma interactions play a central role in a wide range of applications ranging from table-top laser wakefield accelerators~\cite{revacc,eloy} and light sources~\cite{mourou} to laser fusion~\cite{bib:kirkwood_ppcf_2013}. So far, laser plasma research has been geared towards the use of drivers with Gaussian transverse profiles. The use of these laser drivers in laser wakefield acceleration (LWFA), for instance, has lead to remarkable achievements~\cite{lwfa}. Important LWFA experimental results have been reached in the strongly nonlinear bubble or blowout regime~\cite{bib:pukhov_apb_2002,bib:kostyukov_pop_2004,bib:lu_prl_2006}. The blowout provides higher accelerating gradients ($\lesssim1~\mathrm{GeV/cm}$) and efficiencies in comparison to the linear regime. In addition, the blowout regime also leads to linear accelerating and transverse focusing forces, where electrons can accelerate with minimal emittance growth~\cite{bib:tzoufras_prl_2008}. Thus, the non-linear regime of plasma-based wakefield acceleration has the potential to lead to a high quality electron source for science and applications.

Although ideally suited for electron trapping~\cite{bib:kalmykov_prl_2009} and acceleration~\cite{bib:lu_prstab_2006}, the blowout is not adequate for positron acceleration. Instead of bringing positrons towards the axis, the transverse focusing force in the blowout regime defocuses positrons everywhere except in a narrow region where plasma electrons cross the axis at the back of the bubble. Several positron acceleration schemes have then been proposed in order to optimise positron acceleration in plasmas. The nonlinear suck-in regime, for instance, provides positron accelerating and focusing fields similar to the linear/mildly nonlinear regime~\cite{bib:lee_pre_2001}. Hollow plasma channels~\cite{bib:schroeder_pop_2013,bib:kimura_prstab_2011} were also proposed as a mean to accelerate positrons in the linear regime~\cite{bib:esarey_ieee_1996}. Despite these advances, studying new configurations for positron acceleration in the nonlinear regime is important and may have a strong impact in future plasma based linear colliders.

There has been an increasing interest on the interaction between plasmas and lasers with orbital angular momentum (OAM). The production of these states, described by higher order Laguerre-Gaussian modes, is currently well understood~\cite{allen}. Raman and Brillouin backscattering~\cite{mendprl}, and the inverse Faraday effect were then recently examined~\cite{ali} using higher order lasers in plasmas. Linear wakefield excitation by higher order Laguerre-Gaussian laser pulses was also recently investigated~\cite{bib:mendonca_pop_2014}. Nevertheless, the use of lasers with OAM to drive non-linear wakefields for electron and positron acceleration is still unexplored.

In this work we demonstrate analytically and through 3D particle-in-cell (PIC) simulations in OSIRIS~\cite{bib:fonseca_book} that laser pulses with OAM can excite high-gradient positron focusing and accelerating wakefields in the non-linear regime. We show that there are two limiting scenarios: at lower laser intensities, close to the onset of the blowout, the driver excites a donut shaped bubble which can trap and accelerate hollow electron bunches. These results open the way to produce transversely tailored relativistic electron bunches for applications~\cite{bib:stancari_prl_2011}. At higher laser intensities, the inner electron sheath that surrounds the donut wakefield merges on-axis. This results in strongly nonlinear wakefields that can focus and accelerate positron bunches. Moreover, the resulting focusing force can be more than an order of magnitude stronger than in a pure ion channel.

To investigate wakefield excitation by lasers with OAM, we adopt the co-moving frame variables, where $(x,y,\xi=z-ct,t)$, with $(x,y)$ the transverse coordinates, and $t$ and $z$ the time and propagation distance. In addition, we use cylindrical coordinates where $r=\sqrt{x^2+y^2}$ is the distance to the axis and where $\theta$ is the azymuthal angle. The normalised vector potential ($a_{\mathrm{L}}=e A_{\mathrm{L}}/m_e c$) of a laser with OAM at the focus is given by $a_{\mathrm{L}}(r,\xi) = a_0 a_{\|}(\xi) a_{r}(r)$, where $a_0$ is the peak laser vector potential, $a_{\|}(\xi)$ is the longitudinal intensity profile normalised to $a_0$, and $a_{r}(r)$ is the transverse laser profile given by $a_{r}(r)= c_{l,p}(r/w_0)^{|l|} \exp\left(-r^2/w_0^2+il\theta \right) L_p^{|l|}(2 r^2/w_0^2)$, where $w_0$ is the spot-size, $L_p^l$ is a Laguerre polynomial with radial index $p$ and azimuthal index $l$ and $c_{l,p}$ are normalising factors. Unless explicitly stated, we normalise electric fields $\mathbf{E}$ to $m_e c^2\omega_p/e$, magnetic fields $\mathbf{B}$ to $m_e c\omega_p/e$, vector potentials $\mathbf{A}$ to $m_e c/e$, distances $\mathbf{x}$ to $c/\omega_p$, and time $t$ to $\omega_p$. Here $m_e$ and $e$ are the electron mass and charge, $\omega_p$ the plasma frequency, and $c$ the speed of light. We also normalise densities to the background plasma density $n_0$ and velocity $\mathbf{v}$ to $c$.

Linear theory for wake excitation by lasers with OAM has been derived in the limit where the laser central frequency ($\omega_0$) is much higher than the plasma frequency ($\omega_p$). Corresponding accelerating ($E_z$) and focusing ($W_r$) wakefields acting on a relativistic particle are given by $E_z=-(1/4)\int_{-\infty}^{\xi}\sin\left(\xi-\xi^{\prime}\right)\frac{\partial a_{\mathrm{L}}^2(r,\xi^{\prime})}{\partial\xi} \mathrm{d}\xi^{\prime}$~\cite{bib:esarey_ieee_1996,bib:mendonca_pop_2014,bib:michel_prstab_2011}, and by $\partial W_r/\partial\xi=\partial E_z/\partial r$, where $W_r=E_r-B_{\theta}$, and where $E_r$ and $B_{\theta}$ are the transverse electric and azimuthal magnetic fields. For a flat top laser pulse driver with $a_{\|}=\Theta(\xi)-\Theta(\xi-\lambda_p/2)$, where $\Theta(x)$ is the Heaviside step function, and $(l,p)=(1,0)$, these expressions lead to: 
\begin{subequations}
\label{eq:wake_linear}
\begin{align}
E_z & =\frac{a_r^2(r)}{4}\cos(\xi) \label{eq:ez_linear} \\
W_r & =\frac{1}{r}\left(1-\frac{2 r^2}{w_0^2}\right)a_r^2(r)\sin(\xi) \label{eq:wr_linear},
\end{align}
\end{subequations}
for $\xi>\lambda_p/2$. Equation~(\ref{eq:ez_linear}) shows that the amplitude of the wakefield is maximised in the region where $a_r^2(r)$ is maximum, i.e. in a donut shaped region around $r_{\mathrm{m}}\simeq w_0/\sqrt{2}$. In addition, Eq.~(\ref{eq:wake_linear}) shows that focusing and accelerating wakefields overlap by one fourth of the plasma wavelength within the donut shaped region. Although with much lower accelerating gradients, electron and positron focusing can also occur near the axis, around $r=0$. The overlap between focusing $W_r$ and accelerating $E_z$ fields still exists at $r=0$ for one fourth of the plasma wavelength. 

\begin{figure}
\centering\includegraphics[width=.7\columnwidth]{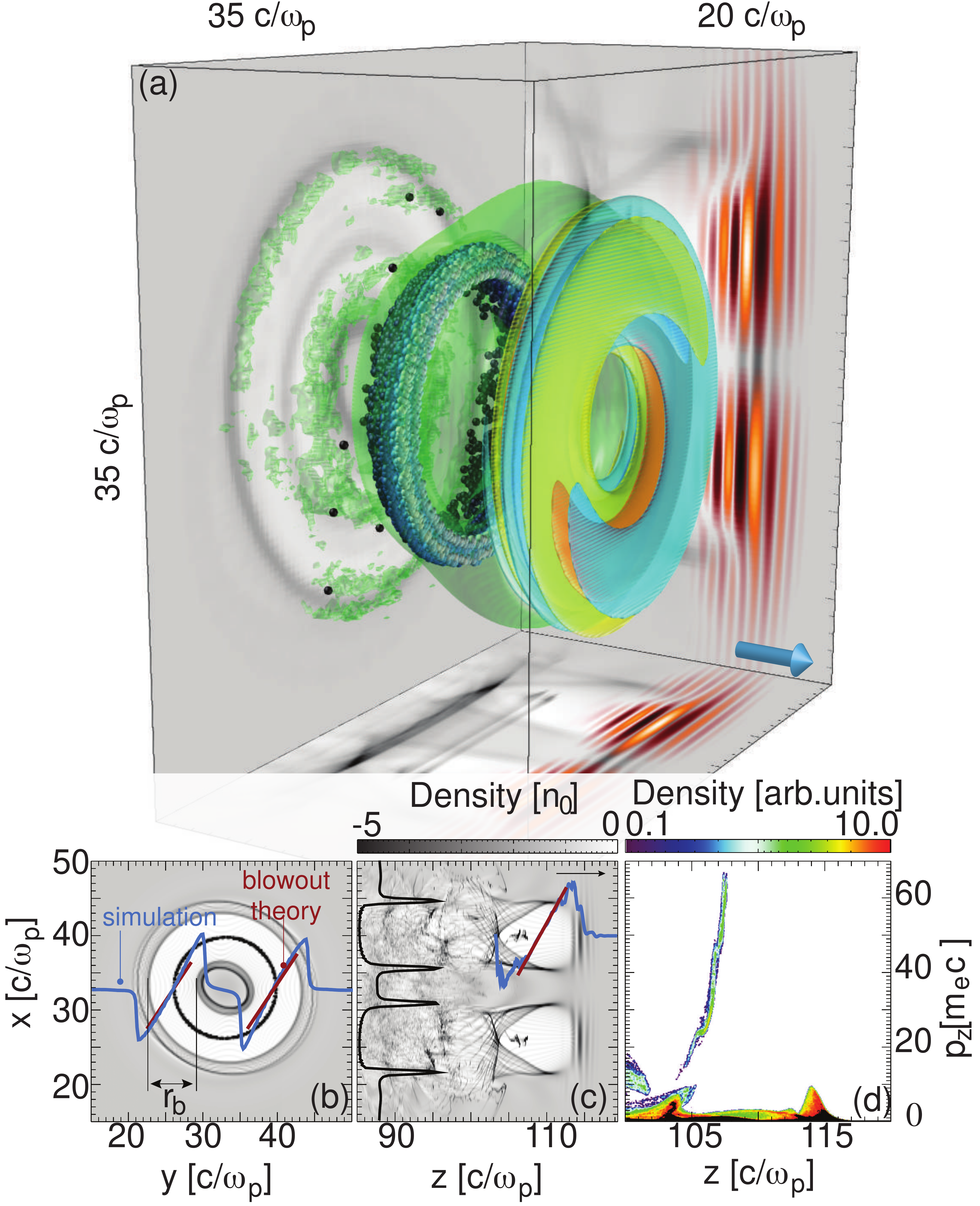}
\caption{OSIRIS simulation result of the self-injection of a hollow electron bunch. (a) green-blue colours are plasma density isosurfaces. White-blue spheres are self-injected electrons coloured according to their energy (higher energy in white and lower energy in blue). Projections show the plasma density in grey and laser fields in orange-red-brown colours. (b) shows a transverse slice of the simulation box (laser propagates outside the plane) superimposed by a transverse lineout of plasma focusing fields took in the region where the blowout radius is maximum. Simulation results are in blue and analytical theory in red.(c) shows a transverse slice of the box. A lineout of the accelerating gradient took at the center of the bubble is shown in blue and theoretical prediction in red. The solid black line is a lineout of the plasma density at $z=110~c/\omega_p$. (d) shows the plasma phase space, where the hollow electron bunch gains up to 35 MeV. Arrows indicate the direction of laser propagation.}
\label{fig:hollow}
\end{figure}

In order to show that both positrons and electrons can focus and accelerate in non-linear donut shaped wakefields,
we start by examining a scenario where the laser intensity is marginally above the threshold for the onset of the blowout regime. Figure~\ref{fig:hollow} then shows simulation results using a laser with $(l,p)=(1,0)$, $a_0=3.2$, $w_0=7~\mathrm{c/\omega_p}$, FWHM duration $\tau = 4/\omega_p$, and $\omega_0/\omega_p=5$. Considering $\lambda_0=2\pi c/\omega_0=800~\mathrm{nm}$ this corresponds to a laser with $\simeq140$ mJ and $w_0=4.5~\mu\mathrm{m}$ with $\tau=10~\mathrm{fs}$ propagating in a plasma with $n_0=6.9\times10^{19}~\mathrm{cm}^{-3}$. The simulation uses a moving window propagating at $c$, with dimensions $20\times62.5\times62.5~(c/\omega_p)^3$, divided into $3000\times325\times325$ cells with $2\times1\times1$ particles per cell.

Figure~\ref{fig:hollow}a shows the electron density, the laser projections, and trapped plasma electrons forming an hollow electron bunch. This result is in stark contrast with self-injection in the spherical blowout regime, which produces cylindrical beams. In addition to their fundamental importance, as they can carry currents which can exceed the Alfven current~\cite{bib:davies_pre_2003}, hollow bunches can also be relevant for applications. For instance, it has been recently shown that they could be used as a compact collimator for proton bunches in conventional accelerators~\cite{bib:stancari_prl_2011}.

Figure~\ref{fig:hollow}b shows a transverse density slice of the donut wakefield. The darker density ring corresponds to the self-injected hollow electron bunch. The lighter density rings, inside and outside of the darker self-injected electron ring, define the donut shaped bubble radially. This structure provides forces that focus the hollow self-injected electron bunch during its acceleration. Focusing forces are well described by the analytical scaling for the spherical blowout, $W_r=(r-r_\mathrm{m})/2$, as shown by Fig.~\ref{fig:hollow}b. The hollow bunch electrons can then perform betatron oscillations around $r_{\mathrm{m}}\simeq w_0/\sqrt{2}$. Corresponding betatron radiation patterns may lead to the generation of hollow x-rays. Similarly to the spherical blowout, hollow self-injection also starts at the back of the donut shaped blowout. This can be seen in Fig.~\ref{fig:hollow}c, which shows the donut wakefield appearing as two symmetric bubbles. The accelerating fields at $r=r_{\mathrm{m}}$ (blue line in Fig.~\ref{fig:hollow}c) also follow the scalings for the spherical blowout, $E_z=\xi/2 (m_e c\omega_p^2/e)$ (red line in Fig.~\ref{fig:hollow}c)~\cite{bib:lu_prl_2006,bib:kostyukov_pop_2004,bib:lu_prstab_2006}.

Since both $E_z$ and $W_r$ are well described by the spherical blowout scaling laws, we can then use them to estimate the maximum energy gain ($\Delta E$) by the hollow bunches. For the parameters of this simulation, $\Delta E \simeq (2/3) m_e c^2 (\omega_0/\omega_p)^2\sqrt{a_0}\simeq 27~\mathrm{MeV}$~\cite{bib:lu_prstab_2006}. This is close to the maximum energies observed in the simulation (Fig.~\ref{fig:hollow}d). Simulations also showed that the laser is self-guided by the donut shaped blowout during electron acceleration, showing that energy is efficiently transferred from the laser to the hollow electron bunch. Therefore, although more refined models can be constructed, these results indicate that spherical blowout theory appears to accurately predict key donut wakefield properties in the blowout regime.

\begin{figure}
\centering\includegraphics[width=\columnwidth]{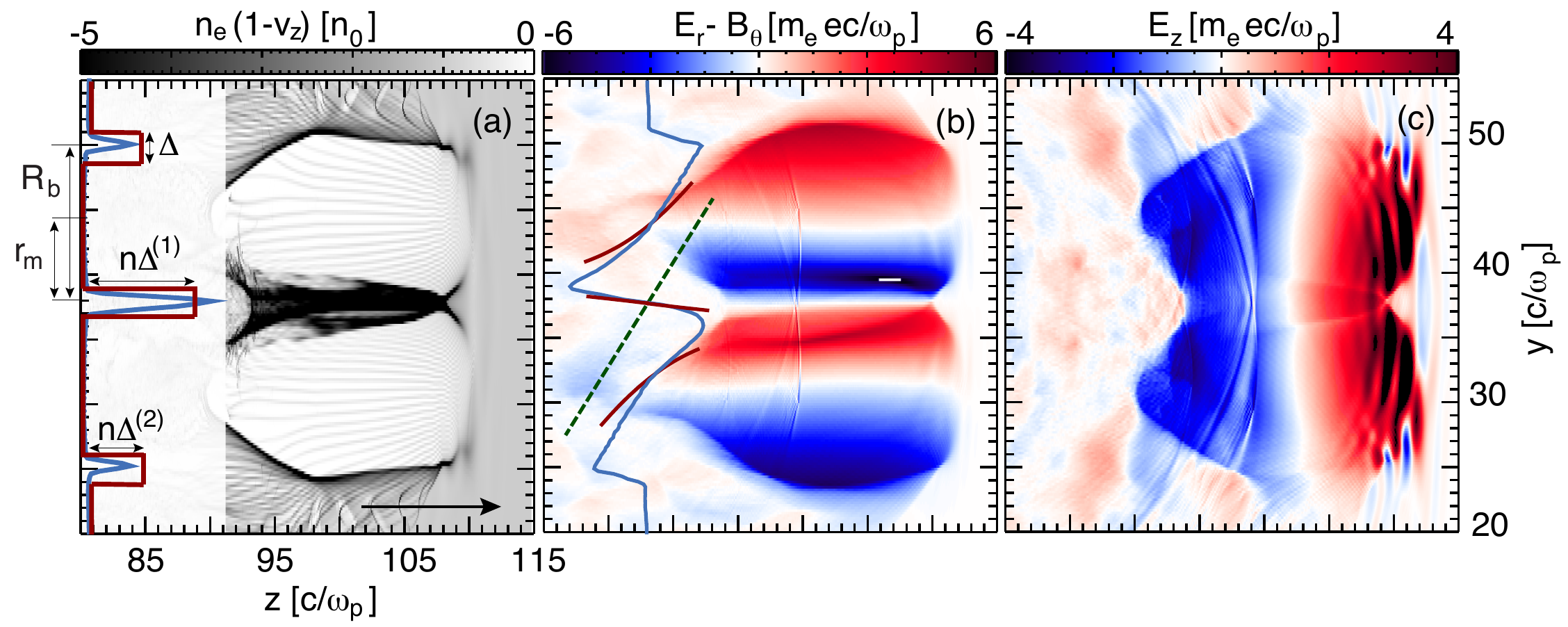}
\caption{OSIRIS simulation result illustrating large amplitude focusing and accelerating fields for positrons in nonlinear regimes. (a) transverse slice of $n_e(1-v_z)$. The blue line is a lineout at $z=115~c/\omega_p$, and the red line is the simplified model used to determine $W_r$. $R_b$ is the distance from the axis to the outer blowout donut electron sheath, $r_\mathrm{m}$ is the distance from the axis to the center of the blowout donut, $\Delta$ is the thickness of the electron sheaths, and $n_{\Delta}^{(1)}$ and $n_{\Delta}^{(2)}$ their corresponding $n_e(1-v_z)$ values. (b) transverse slice of the transverse focusing force. The blue line is a lineout at $z=115~c/\omega_p$, the red line is the theoretical result, and the dashed green line is the theory for the spherical blowout. (c) transverse slice of the accelerating electric field. The arrow indicates the direction of laser propagation.}
\label{fig:positron}
\end{figure}

Donut bubble transverse wakefields are able to focus and accelerate positrons on-axis when the laser radial ponderomotive force is sufficiently intense to push the inner donut electron sheath towards $r=0$. When this occurs, the electron charge density on-axis becomes much higher than the background plasma ion density, which results in a positron focusing field. This conditions is met when the donut blowout radius, $r_b\simeq2\sqrt{a_0}$, matches $r_{\mathrm{m}}= w_0/\sqrt{2}$. The required laser intensity is then given by $a_0 \simeq w_0^2/8$ in our normalised units. We can predict the resulting positron focusing force analytically by recalling that in the blowout regime, $W_r=E_r-B_{\theta}$ is electromagnetic in character due to the presence of strong plasma currents. Under the quasi-static approximation and in the Lorentz Gauge, $E_r = -\partial_r\phi - \partial_{\xi} A_r$ and $B_{\theta} = -\partial_r A_z-\partial_{\xi} A_r$, where $\phi$ is the scalar potential, and where $A_r$ and $A_z$ are the radial and longitudinal plasma (slow varying) vector potentials. As a result, $E_r-B_{\theta} = -\partial_r\phi + \partial_r A_z = -\partial_r (\phi-A_z) = -\partial_r \psi$. The plasma pseudo potential, $\psi\equiv\phi-A_z$, hence fully determines plasma focusing fields, including the contributions from the ion and electron density distribution and plasma currents~\cite{bib:lu_thesis}. In order to obtain an expression for $\psi$, we solve $-\nabla_{\perp}^2\psi = 4\pi \left[n_e\left(1-v_{\|}\right)-1\right]$, or, equivalently, $\psi=\ln(r)\int_0^r r^{\prime} \mathrm{d} r^{\prime} \left\{n_e(r^{\prime})[1-v_z(r^{\prime})]-1\right\} + \int_r^{\infty} \ln(r^{\prime}) r^{\prime} \left\{n_e(r^{\prime})[1-v_z(r^{\prime})]-1\right\}$~\cite{bib:lu_thesis} where $\nabla_{\perp}^2=(1/r)\partial_r(r\partial_r\psi)$ is the transverse Laplacian, and where $n_e\left(1-v_{\|}\right)-1$ is the source term for $\psi$. We consider a simplified model for the source term $n_e\left(1-v_z\right)$ shown in Fig.~\ref{fig:positron}a, where the blue line represents the simulation $n_e(1-v_z)$, and the red line, the simplified model for the calculation. Values for $n_e\left(1-v_{\|}\right)$ at $r=0$ ($n_{\Delta}^{(1)}$) and at the bubble wall ($n_{\Delta}^{(2)}$) can be derived by noticing that $\mathrm{d}/\mathrm{d}\xi \left\{\int[n_e(1-v_{\|})]\mathrm{d}\mathbf{x}_{\perp}\right\}=0$, i.e. $n_e(1-v_z)$ is conserved in each transverse slice. Assuming further that $\int_0^{R_b/2}[n_e(1-v_{\|})-1]\mathrm{d}\mathbf{x}_{\perp} = \int_{R_b/2}^{R_b}[n_e(1-v_{\|})-1]\mathrm{d}\mathbf{x}_{\perp}=0$ , then $n_{\Delta}^{(1)}=R_b^2/(4\Delta^2)$ and $n_{\Delta}^{(2)} = [(R_b+\Delta)^2-(R_b/2)^2]/[(R_b+\Delta)^2-R_b^2]$, where $\Delta$ is the thickness of the electron layers defining the donut blowout (see Fig.~\ref{fig:positron}a). In the relativistic blowout regime $R_b\gg 1$ and $\Delta\ll R_b$ such that:
\begin{equation}
\label{eq:psi1}
\psi = \frac{1}{8}\left[2(R_b^2-r^2)+R_b^2 \ln\left(\frac{r}{R_b}\right)\right] + \frac{3 R_b^2 \alpha}{16},
\end{equation} 
for $\Delta < r < R_b$, where $R_b=2 r_b$, and where $\alpha \equiv \Delta/R_b\ll 1$. In addition,
\begin{equation}
\label{eq:psi2}
\psi = \frac{r^2}{16 \alpha^2}+\frac{1}{16}\left\{8 r^2[\ln(r)-1]+R_b^2[3+2\ln(\alpha)]\right\},
\end{equation} 
for $r<\Delta<R_b$.
The focusing force is thus:
\begin{equation}
\label{eq:foc1}
W_r = \frac{r}{2}-\frac{R_b^2}{8 r},
\end{equation} 
for $\Delta < r < R_b$. The first term in Eq.~(\ref{eq:foc1}) is due the ion column and the second to the thin on-axis electron layer. In addition:
\begin{equation}
\label{eq:foc2}
W_r = \frac{r}{2}\left[1-\frac{1}{4\alpha^2}-\ln(r)\right],
\end{equation} 
for $r<\Delta<R_b$. The first term in Eq.~(\ref{eq:foc2}) is due to the ion column and the second and third due to the on-axis electron layer.

Equation~(\ref{eq:foc2}) reveals that the positron focusing fields in donut shaped bubbles can be much higher than the electron focusing fields in spherical bubbles, as long as $r<\Delta$. According to Eq.~(\ref{eq:foc2}) for $\alpha\ll 1$, the on-axis positron focusing force is $W_r^{e^+}\simeq r/(8\alpha^2)$, whereas for a pure ion spherical bubble $W_r^{e^-}=1/2$. Hence $W_r^{e^+}/W_r^{e^-}=1/(4\alpha^2)$. For $\alpha \simeq 1/10$, $W_r^{e^+}$ would then be 50$\times$ stronger than $W_r^{e^-}$. This may lead to the emission of higher frequency x-rays by positron bunches in the donut blowout than by electrons in a pure ion spherical bubble~\cite{bib:kostyukov_pop_2003}. It is important to note that the width of the positron focusing region extends well beyond the width of the on-axis electron layer, lasting from $0<r<R_b/2$. Moreover, we note that although they may vary due to the dynamics of the inner donut sheath, positron accelerating fields are nevertheless similar to those of the spherical blowout. Positron acceleration can therefore occur for the first half of the donut.

We confirmed these findings in 3D PIC simulations. Figure~\ref{fig:positron} was obtained using identical simulation parameters to Fig.~\ref{fig:hollow} except for the laser $a_0$. The onset of positron focusing and acceleration fields on axis for the laser spot-size of the simulation shown in Fig.~\ref{fig:hollow} ($w_0=7 c/\omega_p$) is $a_0\simeq6.12$. Figure~\ref{fig:positron}, in which $a_0=8>6.12$, then confirms that $r_b>w_0$ and that the inner donut bubble electron sheath merges on axis. These parameters correspond to a laser pulse energy of $\simeq 850~\mathrm{mJ}$. Figure~\ref{fig:positron}a shows that the motion of the boundary $R_b(\xi)$ resembles that of a spherical bubble. Figure~\ref{fig:positron}b confirms that the donut bubble focusing fields (blue line) are much higher than the focusing fields of a spherical bubble (dashed green line), and that in this case $W_r^{e^+}/W_r^{e^-}=14$. In addition, Fig.~\ref{fig:positron}b also shows that simulation results for $W_r$ are in good agreement with the predictions of Eq.~(\ref{eq:foc1}) and (\ref{eq:foc2}) using $R_b=11.8$ and $\Delta=1.5$ (red line). Values for $R_b$ and $\Delta$ were retrieved directly from Fig.~\ref{fig:positron}a at $z=113.5~c/\omega_p$. Figure~\ref{fig:positron}c shows very large positron accelerating gradients that overlap with positron focusing wakefields for the first half of the bubble. The peak positron accelerating fields in Fig.~\ref{fig:positron}c is $15~\mathrm{GV/cm}$, being higher than the peak electron accelerating field at the back of the donut.

\begin{figure}
\centering\includegraphics[width=\columnwidth]{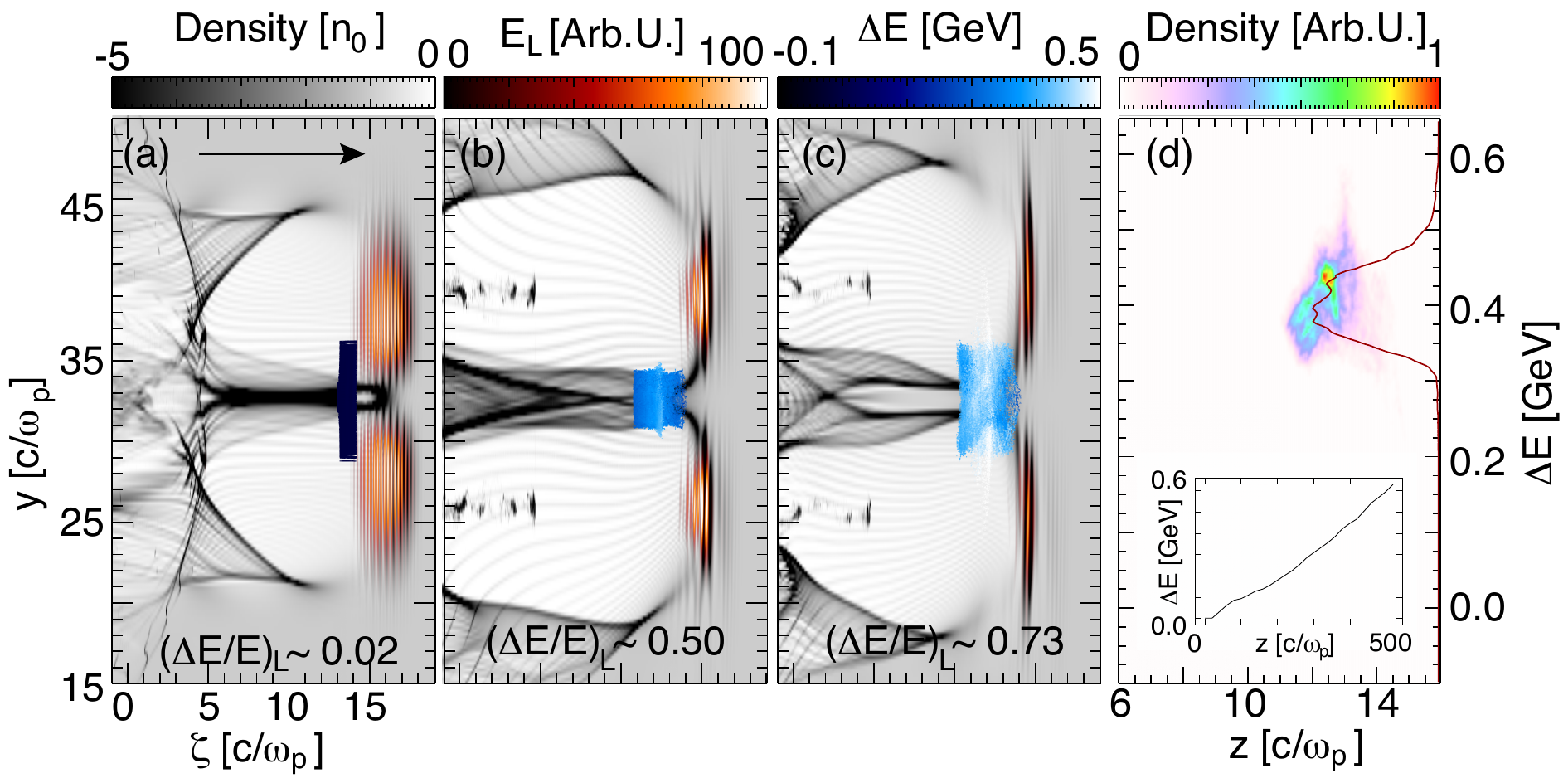}
\caption{OSIRIS simulation result illustrating positron acceleration in the wake driven by a laser pulse with OAM. (a)-(c) plasma density (gray), laser pulse electric field (orange) and witness positron bunch (blue) coloured according to energy gain at $z=40 c/\omega_p$, $z=400~c/\omega_p$ and $z=518~c/\omega_p$. The ratio $(\Delta E/E)_L$ indicates the fraction of the laser energy that has been depleted. The arrow indicates the laser propagation direction. (d) Longitudinal phase-space of the witness positron bunch and energy spectrum (red line) at $z=518~c/\omega_p$. The witness positron bunch was initially injected with $\gamma=200$. The inset shows the maximum energy of the positron bunch as a function of the propagation distance.}
\label{fig:acceleration}
\end{figure}

We also performed simulations to demonstrate the acceleration of a witness positron bunch in the donut shaped wakefields. Figure~\ref{fig:acceleration} shows an OSIRIS 3D simulation result using a laser pulse with $(l,p)=(1,0)$, $w_0=6.6 c/\omega_p$, $a_0=6.8$, $\omega_0/\omega_p=15$ and $\tau=3/\omega_p$ (shorter than in the previous simulations to reduce direct interaction between witness positron bunch and laser fields). We placed a flat-top witness positron bunch with a length of $1~c/\omega_p$ and transverse radius $3~c/\omega_p$ on axis, in regions where the initial positron accelerating fields were close to maximum. The witness bunch was injected with $\gamma=200$ (ensuring that all positrons are trapped) with zero energy spread and zero emittance. The charge of the witness bunch is sufficiently low to avoid beam loading. Thus, positrons effectively behave as test particles. The simulation box is $20\times65\times65 (c/\omega_p)^3$ divided into $3000\times 325\times 325$ cells with 2 plasma electrons and positrons per cell. For a laser pulse with $\lambda_0=800~\mathrm{nm}$ this corresponds to a laser with $w_0=12.6~\mu\mathrm{m}$, $\tau = 19~\mathrm{fs}$ and and energy of 2~J propagating in a plasma with $n_0=7.7\times10^{18}~\mathrm{cm}^{-3}$. For this plasma density the witness bunch is $3.84~\mu\mathrm{m}$ long and $5.76~\mu\mathrm{m}$ wide.

Figure~\ref{fig:acceleration}a-b shows that the donut shaped bubble can guide the laser pulse. We found in simulations that the self-guiding condition for the spherical blowout regime, given by $w_0 \simeq r_b \simeq 2\sqrt{a_0}$, can also be used as a guide for the self-guided propagation of a donut shaped laser. This can be attributed to the fact that the donut bubble refracting index gradient, which determines the laser dynamics, is identical to that of spherical bubbles. The simulation of Fig.~\ref{fig:acceleration} then used $w_0 = 6.6$, which is close to $2\sqrt{a_0}\simeq 5.2$. Although the on-axis electron sheath oscillates during the laser propagation for these parameters, the plasma wave still provides positron accelerating and focusing fields. Figures~\ref{fig:acceleration}a-c show that the laser can drive a stable wakefield that ensures positron acceleration until the laser energy is a small fraction of its initial energy. This is confirmed by the inset of Fig.~\ref{fig:acceleration}d, which gives the evolution of the maximum positron energy as a function of the propagation distance, revealing constant acceleration gradients along the propagation. Resulting normalised average accelerating gradient is $E_{\mathrm{accel}}\simeq 1.5$, confirming that positrons accelerated with high acceleration gradients much larger than those of the linear regime for which $E_{\mathrm{accel}}\ll 1$. After $z=518~c/\omega_p$, the mean positron bunch energy gain is $\Delta E\simeq400$ MeV (Fig.~\ref{fig:acceleration}d). We note that the mean energy gain of self-injected electrons is similar to the positron energy gain ($\Delta E\simeq 400~\mathrm{MeV}$). Simulations also show that, in agreement with analytical predictions, the transverse donut shaped wakefields focus the positron bunch even though its initial radius is much wider than the on-axis electron sheath~\cite{bib:thankyou}.

We note that simulations using hollow electron bunch drivers showed similar initial positron focusing and acceleration regions in nonlinear regimes. These simulations then indicate that our results are determined by the intensity profile of the driver, being independent of its phase content.

In conclusion, we investigated self-injection of hollow electron beams and positron acceleration in donut shaped wakefields driven by higher order Laguerre-Gaussian laser pulses with OAM in the nonlinear regime analytically and with 3D PIC simulations. We showed that a laser with OAM can be self-guided by the donut shaped blowout region. The onset of positron focusing and accelerating fields occurs when the electron sheath at the inner wall of donut shaped blowout merges on axis. Resulting positron focusing could lead to enhanced betatron x-ray radiation emission regimes in comparison to electrons in the spherical blowout regime. More detailed understanding of externally-guided and self-guided regimes will be important to improve the stability of positron and electron acceleration.  

\acknowledgements
This work has been partially supported by FCT (Portugal) through grant EXPL/FIS-PLA/0834/2012. We acknowledge PRACE for access to resources on SuperMUC (Leibniz Research Center). We also acknowledge useful discussions with Prof. Lu\'is O. Silva.


\begin{thebibliography}{99}
\bibitem{bib:thankyou} The authors would like to thank the anonymous referee for the suggestion to include a 3D simulation to demonstrate positron bunch acceleration in the donut blowout regime.
\end{thebibliography}
\end{document}